\documentclass[a4paper]{article}

\usepackage{authblk}
\usepackage{hyperref}
\usepackage{amsfonts,amssymb,amsmath,mathtools} 
\usepackage{graphicx}
\usepackage[symbol*]{footmisc}
\usepackage{xcolor}

\usepackage{caption}
\usepackage{subcaption}
\usepackage[absolute,overlay]{textpos}

\newtheorem{proposition}{Proposition}[section]
\newtheorem{theorem}[proposition]{Theorem}

\newtheorem{definition}[proposition]{Definition}
\newcommand {\proof} {\par\textit{Proof}. \ignorespaces}
\newcommand {\eproof}
      {\null\hfill{\large$\Box$}}
\newenvironment{prf}{\proof}{\eproof}

\newcommand{\vektor}[1]{{\boldsymbol{#1}}}

\newcommand{\vx}{\mathbf{x}}
\newcommand{\bi}{\mathbf{i}}
\newcommand{\vy}{\vektor{y}}
\newcommand{\vz}{\vektor{z}}
\newcommand{\vPhi}{\vektor{\Phi}}

\newcommand{\coveval}{\beta}

\DefineFNsymbols*{lamport}{*{\textsuperscript{\textdagger}}\ddagger\S\P\|%
{**}{\dagger\dagger}{\ddagger\ddagger}}

\setfnsymbol{lamport}

\begin{document}

\title{Quasi-Monte Carlo methods for lattice systems:\\a first look}

\renewcommand{\thefootnote}{\fnsymbol{footnote}}
\renewcommand{\Affilfont}{\itshape\small}

\author[a,b]{K.~Jansen}
\author[c]{H.~Leovey}
\author[a,d]{A.~Ammon}
\author[c]{A.~Griewank}
\author[d]{M.~M{\"u}ller-Preussker}

\affil[a]{NIC, DESY Zeuthen, Platanenallee 6, D-15738 Zeuthen, Germany}
\affil[b]{Department of Physics, University of Cyprus, P.O.Box 20537, 1678 Nicosia, Cyprus}
\affil[c]{Institut f\"ur Mathematik, Humboldt-Universit\"at zu Berlin, Unter den Linden 6, D-10099 Berlin}
\affil[d]{Institut f\"ur Physik, Humboldt-Universit\"at zu Berlin, Newtonstr. 15, D-12489 Berlin}

\maketitle

\begin{textblock}{15}(29,2)
\setlength{\parindent}{0cm}
HU-EP-13/05\\
SFB/CPP-13-13\\
DESY 13-022
\end{textblock}

\begin{abstract}
We investigate the applicability of 
Quasi-Monte Carlo methods to Euclidean lattice systems 
for quantum mechanics in order to improve the asymptotic 
error behavior of observables for such theories. In most cases the error of an observable 
calculated by averaging over random observations generated from an ordinary 
Markov chain Monte Carlo 
simulation behaves like $N^{-1/2}$, where $N$ is the number of observations. By means 
of Quasi-Monte Carlo methods it is possible to improve this behavior for certain problems 
to $N^{-1}$, or even further if the problems are regular enough. 
We adapted and applied this approach  to simple systems like the 
quantum harmonic and anharmonic oscillator and verified an improved error scaling. 
\end{abstract}

\section{Introduction}

Markov chain-Monte Carlo (Mc-MC) techniques are commonly the method of choice
for the numerical evaluation of partition functions 
in statistical physics or path integrals in Euclidean time
for models in high energy physics. The reason is that they are based on 
importance sampling and hence select the integration points automatically according 
to the corresponding weight in the integrand. 
Many algorithms have been developed to implement a Mc-MC, starting from 
the Metropolis algorithm, heatbath and over-relaxation to cluster and hybrid Monte Carlo algorithms, 
see e.g. refs.~\cite{Gattringer:2010zz,Montvay:1994cy}. 
In this way, simulations of demanding 
4-dimensional quantum field theories became possible and, in fact, were carried 
out very successfully to e.g. compute the low-lying hadron spectrum 
\cite{Fodor:2012gf}
or deriving 
bounds for the Higgs boson mass \cite{Bulava:2012rb}. 

The drawback of Mc-MC is that it estimates the desired quantity stochastically and hence 
the results are affected by a statistical error which sometimes needs very long 
and computer time extensive samplings. Quantitatively, this sampling error behaves as
$N^{-1/2}$ for a (thermalized) sample size of $N$.  
This error scaling behaviour is often a real stumbling block in such Mc-MC
simulations. If we consider lattice quantum chromodynamics as a typical system 
for Mc-MC calculations in high energy physics, then due to this 
error scaling and the very high computational demand of 
these simulations, it is often impossible to significantly decrease the error 
to the targeted precision.   
It would therefore be very desirable to have Monte Carlo methods available that 
possibly show a better error scaling. 

Such methods in fact exist in form of Quasi-Monte Carlo 
(QMC)
techniques 
\cite{L'Ecuyer01},\cite{KSS_Review12}, where it is known that 
the error scaling can be improved to an $N^{-1}$, 
or even more if the problem exhibits enough structure (smoothness, periodicity). 
QMC methods have been analyzed theoretically very 
thoroughly and comprehensively, and many successful 
applications in mathematical finance based on high-dimensional Gaussian integrals 
have been studied in the last two decades \cite{Glasserman},\cite{jaeckel02}.
On the other hand, to the best of
our knowledge, QMC methods were never tested successfully in high-dimensional models that are relevant for high energy physics. 

%
In this paper we therefore want to perform a very first step towards the 
very challenging goal of applying QMC methods to generic field theories by
looking at the non-trivial case of the anharmonic quantum mechanical 
oscillator discretized on a finite Euclidean time lattice and evaluated in the 
corresponding path integral formulation \cite{Creutz_and_Freedman}. 
For the case of the anharmonic oscillator
the system is not Gaussian anymore and a successful application of 
QMC methods would be a first non-trivial test. 
Of course, even if such a test is successful, there is a long way to address 
eventually 4-dimensional quantum field theories, but a proof of concept would certainly 
open the promising road to attack field theories in the future. 

We will start our discussion with a description of the harmonic oscillator 
in its time-discretized path integral formulation. Here the problem is fully
Gaussian and the application of adequate QMC methods should lead to an improved 
error scaling, an expectation that we will see to be fulfilled.  
Nevertheless, the harmonic oscillator example can serve well to explain
how QMC methods work and how an improved error scaling behaviour is realized. 

We will then proceed to look at the anharmonic oscillator and we will demonstrate 
that also in this case QMC leads to an improved error scaling, although the obtained 
rate of convergence is still not optimal and leaves space for improvements. 
We consider this, nevertheless, to be a very promising 
non-trivial result which bears the potential that also other models in quantum 
mechanics, e.g. the topological quantum mechanical action of ref.~\cite{Bietenholz:2010xg} 
and even field theories can be evaluated by QMC methods.   
For a first account of our studies, we refer to the proceedings contribution 
of ref.~\cite{Jansen:2012gj}.

\section{Quantum Mechanical Harmonic and Anharmonic Oscillator}
In this section we will discuss the basic steps for the quantization of the theory 
in the path integral approach and the discretization on a time lattice.
The first step is the construction of the Lagrangian (resp. the action) of the 
corresponding classical mechanical system for a given path $x(t)$ of a particle 
with mass $M_0$. For a numerically stable evaluation of the path integral it is 
essential to pass on to Euclidean time. In this case the Lagrangian $L$ and the 
action $S$ is given by:
\begin{align}
\label{eq:Lagrangian}
  L(x,t) &= \frac{M_0}{2}
  \left(\frac{d x}{dt}\right)^2 + V(x) \\
  S(x) &= \int_0^T \, L(x,t) \; dt  .
\end{align}
Depending on the scenario (harmonic or anharmonic oscillator) the potential $V(x)$ 
consists of two parts 
\begin{equation}
V(x) =  \underbrace{\frac{\mu^2}{2} x^2}_\text{harmonic part} + 
\underbrace{\lambda \, x^4}_\text{anharmonic part} \;,
\end{equation}
such that the parameter $\lambda$ controls the anharmonic part of the theory. 
It should also be mentioned that in the anharmonic case the parameter 
$\mu^2$ can take on negative values, leading then to a double well potential.

The next step is to discretize time into equidistant time slices with a 
spacing of $a$. The path is then only defined on the time slices:
\begin{align}
  t  & \rightarrow t_i = (i-1) \cdot a \quad i =  1 \ldots d \\
 x(t) & \rightarrow x_i = x(t_i) \; .
\end{align}
On the lattice the derivative with respect to the time appearing in 
\eqref{eq:Lagrangian} (first term) will be replaced by the forward finite difference 
$\nabla x_i = \frac{1}{a} ( x_{i+1} - x_i )$. The choice of the lattice derivative 
is not unique and requires special care, particularly if one considers more 
complicated models like lattice QCD. But in \cite{Creutz_and_Freedman} it was 
shown that the lattice derivative chosen here permits a well defined continuum 
limit. Putting all the ingredients together, we can write down the lattice action 
for the (an)harmonic oscillator
\begin{equation}
S^\text{latt}(\vx) = a \sum_{i=1}^{ d } 
           \left(\frac{M_0}{2} \left( \nabla x_i \right)^2 + V(x_i)\right) \; .
\end{equation}
For the path a cyclic boundary condition $x_{d+1} = x_1$ can be assumed.
In the following the superscript ``latt'' will be dropped, as we will only 
refer to the lattice action from now on.
The expectation value of an observable $O$ of the quantized theory expressed 
in terms of the path integral reads as follows:
\begin{equation}\label{sec:Plain:OBS}
\left\langle O(\vx) \right\rangle \, = \, \frac{\int_{\mathbb{R}^d} O(\vx)
e^{-S(\vx)} d x_1...d x_d }{\int_{\mathbb{R}^d} e^{-S(\vx)} d x_1...d x_d }\;.
\end{equation}
This expression is suitable for a numerical evaluation of certain quantities of 
the underlying theory. Up to now only Monte Carlo methods are known to give 
reliable results for dimensions $d \gg 10$. One type of such methods, often 
used in physics, is the Markov chain-Monte Carlo approach mostly applying 
the weight $\propto e^{-S(\vx)}$ for sampling paths $\{x_i\}$ (so-called 
``importance sampling''). 
In the next sections, we will provide a summary of the mathematical 
results for 
QMC methods (and their randomizations), and particularly recapitulate 
in a rather mathematical language the strict error scaling bounds 
for this methods. The reader more interested directly in the results 
may move to section 7 directly. 

\section{Direct Monte Carlo and Quasi--Monte Carlo methods}\label{sec:Plain:Int}
We provide in this and the following sections 3-7 some mathematical background of QMC methods. The reader who is more interested in our results, may proceed directly to section 8.
In many practical applications one is interested in calculating quotients of the form 
\eqref{sec:Plain:OBS} where the action $S(.)$ and the observables $O(.)$ are 
usually smooth functions in high dimensions. 
In some special situations where one would like to 
deal with integrands of moderately high dimensions, one possible
approach is to consider estimators $\hat{I}_1$,$\hat{I}_2$ for the integrals 
$I_1$,$I_2$ in the numerator and in the denominator of \eqref{sec:Plain:OBS} separately, 
and then take $\hat{I}_1/\hat{I}_2$ as an estimation of $\left\langle O(\vx) \right\rangle$. 
Another possible approach one can consider is given by the so-called weighted 
uniform sampling (WUS) estimator, analyzed in \cite{PowellSwann66}.
In the latter case, one takes a joint estimator for the total quantity 
$\left\langle O(\vx) \right\rangle$, using a single direct sampling method. 
We will show some characteristics of the WUS estimator in section \ref{sec:WUS}, and we will 
refer from now on to the latter two approaches as \textit{direct} sampling methods 
for estimating \eqref{sec:Plain:OBS}.
In many interesting examples, we encounter the case were the action $S(.)$ 
and the observable $O(.)$ lead to integrals $I_1$,$I_2$ of Gaussian type. 
Then the integrals $I_1$,$I_2$ can be written in the form 
\[
I_i\:=\: \frac{1}{(2\pi)^{d/2} \sqrt{\det(C)}} 
        \int_{\mathbb{R}^d}g_i(\vx)e^{-\frac{1}{2} \vx^\top C^{-1} \vx} d\vx, 
	\quad
	\vx=(x_1,\dots,x_d), \; i=1,2 \quad ,
\]
where $C$ denotes the covariance matrix of the Gaussian density function. 
A transformation to the unit cube in $\mathbb{R}^d$ can be applied such 
that the corresponding integrals take the form 
\begin{equation}\label{gen:expected_v}
I
\:=\:  \int_{[0,1]^d}g(A \vPhi^{-1}(\vz))d\vz
\:=\:  \int_{[0,1]^d}f(\vz)d\vz
\:=\: I_{[0,1]^d}(f)
	,\quad
	\vz=(z_1,\dots,z_d)\,.
\end{equation}
Here $AA^\top=C$ is some symmetric factorization of the covariance matrix, 
and $\vPhi^{-1}(\vz):=(\Phi^{-1}(z_1),\dots,\Phi^{-1}(z_d))^\top$, 
where $\Phi^{-1}({\cdot})$ represents 
the inverse of the normal cumulative distribution function $\Phi({\cdot})$.\\
In the classical direct Monte--Carlo (MC) approach one tries
to estimate \eqref{gen:expected_v} 
by generating samples pseudo-randomly. One starts with a finite sequence of 
independent identically distributed (i.i.d.) samples $P_N=\{\vz_1,\dots,\vz_N\}$, 
where the points $\vz_j, \; 1\le j \le N$, 
have been generated from the uniform distribution in $[0,1]^d$. 
Then, the quadrature rule is fixed by taking the average of the function evaluations for $f$ 
\[
Q_N:= \frac{1}{N} \sum_{j=1}^{N} f(\vz_j),
\]   
as an approximation of the desired integral $\int_{[0,1]^d} f(\vz) \; d \vz$. 
The resulting estimator $\hat{Q}_N$ is unbiased. The integration error 
can be approximated via the central limit theorem, given that $f$ belongs to $L_2([0,1]^d)$. 
The variance of the estimator $\hat{Q}_N$ is given by
\[
\frac{\sigma^2}{N}=\frac{1}{N}\left( \int_{[0,1]^d} f^2(\vz) \; d\vz
  - \left(\int_{[0,1]^d} f(\vz) \; d\vz \right)^2 \right).
\]    
As measured by its standard deviation from zero  
the integration error associated with the 
MC approach is then of order $O(N^{-1/2})$.
The quality of the MC samples relies on the selected pseudo--random 
number generators of uniform samples, here we use the \textit{Mersenne Twister} generator from Matsumoto and Nishimura (see \cite{Matsumoto98}).
MC is in general a very reliable tool in high--dimensional integration, 
but the order of convergence is in fact rather poor. 

In contrast, QMC methods generates deterministically 
point sets that are more regularly distributed than 
the pseudo--random points from MC (see \cite{L'Ecuyer01}, \cite{Novak_and_Wozniakowski2}, 
\cite{DiPi10}, \cite{KSS_Review12}).  
Typical examples of QMC are shifted 
lattice rules and low-discrepancy sequences.   
In order to give a short introduction to the subject, 
we define now the classical notion of discrepancy 
of a finite sequence of points $P_N$ in $[0,1)^d$.    
Given $P_N=\{\vz_{1},\dots,
\vz_{N}\}$ a set of points in $ [0,1)^{d}$, 
and a nonempty family $\mathbb{I}$ of Lebesgue-measurable 
sets in $[0,1)^{d}$, we define the classical discrepancy function by
\[D(\mathbb{I};P_N) := \sup_{B \in \mathbb{I}}\left|\frac{\sum_{i=1}^{N}\:
c_{B}(\vz_{i})}{N}-\lambda_{d}(B)\right|,\]
where $c_{B}$ is the characteristic function of $B$, and $\lambda_{d}$ is the Lebesgue measure in $\mathbb{R}^d$.
This allows us to define the so-called \textit{star discrepancy}

\begin{definition}
We define the \textit{star discrepancy} $D^{\star}(P_N)$ of the point set $P_N$ 
by $D^{\star}(P_N):=D (\mathbb{I};P_N)$, where $\mathbb{I}$ is the family of all 
sub-intervals of the form $\prod_{i=1}^{d}[0,u_{i})$,
with $u_{i}\ge 0, \; 1\le i \le d$.
\end{definition}
The \textit{star discrepancy} can be considered as a measure of the worst difference 
between the uniform distribution and the sampled distribution in $ [0,1)^{d}$ 
attributed to the point set $P_N$. 
The usual way to analyze QMC as a deterministic method 
is by choosing a class of integrand functions $F$, and a 
measure of discrepancy $D(P_N)$ for the point sets $P_N$.
Then, the deterministic integration error is usually given in the form
\[
|Q_N -\int_{[0,1]^d} f(\vz) \; d \vz | \;\; \le \; D(P_N) V(f), 
\]
where $V(f)$ measures a particular variation of the 
function $f \in F$. A classical particular error bound in this form is 
the famous Koksma--Hlawka inequality, 
where $D(P_N)$ is taken to be the \textit{star discrepancy} 
of the point set $P_N$, and $V(f)$ is the variation 
in the sense of Hardy and Krause of $f$. 

In the context of QMC, a sequence of points $\vz_{1},\vz_{2},...$ in $ [0,1)^{d}$ 
is called a low-discrepancy sequence 
if 
\begin{equation}\label{O_disc}
D^{\star}(\{\vz_{1},\dots,
\vz_{N}\})=O(N^{-1}(\log(N))^{d}). 
\end{equation}
An important part of QMC constructions satisfying this asymptotic bound are known under the name of $(t,d)$-sequences and 
will be discussed in more detail in section \ref{t_d_seq}. 
For moderate values of $N$, the influence of the logarithmic term in \eqref{O_disc} usually can not be ignored 
(see $5.7$ and $5.8$ in \cite{Caf_98}), 
because the term $N^{-1}(\log(N))^{d}$ grows until $N>2^d$. 
This normally prevents a straightforward use of low-discrepancy sequences in practical situations with very large 
dimensions. The latter situations are typically very high-dimensional integration problems where 
all variables and interactions between variables are equally important. 

The practical success of QMC sequences in very high-dimensional integration problems with enough smoothness 
usually relies in an appropriate combination with an \textit{effective-dimension} reduction transformation. 
We will discuss the analysis of effective-dimensions more in detail in \ref{eff_sen}.
Well investigated settings for the integration error analysis of functions exhibiting a concentration of 
importance in few variables or groups of few variables are the so-called
\textit{weighted reproducing kernel Hilbert spaces} (see \cite{Novak_and_Wozniakowski2}), 
which will be considered briefly in the following section.\\

\subsection{Quasi--Monte Carlo errors and complexity}\label{erros_complex}
For error analysis of QMC methods, 
there are certain reproducing kernel Hilbert spaces $\mathbb{F}_{d}$ of functions
$f:[0,1]^{d}\to\mathbb{R}$ that are particularly useful (see \cite{Hick98}).
Let us denote now with 
$\langle\cdot,\cdot\rangle$ and $\|\cdot\|$ the inner product
and norm in $\mathbb{F}_{d}$. 
A reproducing kernel is a function $K:[0,1]^{d}\times[0,1]^{d}\to\mathbb{R}$ satisfying 
the properties 
\begin{enumerate}
\item $K(\cdot,\vy)\in\mathbb{F}_{d}$ for each $\vy\in[0,1]^{d}$
\item $\langle f,K(\cdot,\vy)\rangle=f(\vy)$ for
each $\vy\in[0,1]^{d}$ and $f\in\mathbb{F}_{d}$ 
\end{enumerate} 
If the integral
\[
I(f)=\int_{[0,1]^{d}}f(\vz)d\vz
\]
is a continuous functional on the space $\mathbb{F}_{d}$, 
then the worst case quadrature error 
\[
e_{N}(\mathbb{F}_{d}):=\sup_{f\in\mathbb{F}_{d}\,,\|f\|\le 1}|
I(f)-Q_{N}(f)| 
\]
for point sets  $P_N=\{\vz_{1},\dots,\vz_{N}\}$ and QMC algorithms
for the space $\mathbb{F}_{d}$ can be given by 
\[
e_{N}(\mathbb{F}_{d})=\sup_{\|f\|\le 1}|\langle
f,h_{N}\rangle|=\|h_{N}\|
\]
for some $h_{N}\in\mathbb{F}_{d}$ due to Riesz' representation theorem. In this case, the {\em
representer} $h_{N}$ of the quadrature error is given 
explicitly in terms of the kernel by 
\[
h_{N}(\vz)=\int_{[0,1]^{d}}K(\vz,\vy)d\vy -
\frac{1}{N}\sum_{i=1}^{N}K(\vz,\vz_{i}),\quad \forall
\vz\in[0,1]^{d}.
\]
Tensor product reproducing kernel Hilbert spaces are of 
particular interest, since the multivariate kernel results as the 
product of the underlying univariate kernels.
In QMC error analysis, the weighted (anchored) 
tensor product Sobolev space introduced in 
\cite{SlWo98} is often considered
\[
\mathbb{F}_{d}
=\bigotimes_{i=1}^{d}W_{2}^{1}([0,1]),  
\]
also denoted with $\mathbb{F}_{d}=W_{2,{\rm mix}}^{(1,\ldots,1)}([0,1]^{d})$, 
where $W_{2}^{1}([0,1])$ is the Sobolev space of absolutely continuous functions on $[0,1]$ with 
first order derivatives in $L_2([0,1])$. 
The weighted norm $\|f\|_{\gamma}^{2}=\langle
f,f\rangle_{\gamma}$ results from the inner product  
\begin{equation}\label{sec:Error:innerprod}
\langle f,g\rangle_{\gamma}=\sum_{u\subseteq\{1,\ldots,d\}}
\prod_{j\in u}\gamma_{j}^{-1}
\int_{[0,1]^{|u|}}\frac{\partial^{|u|}}{\partial
\vz_{u}}f(\vz_{u},\mathbf{1})\frac{\partial^{|u|}}{\partial
\vz_{u}}g(\vz_{u},\mathbf{1})d \vz_{u},
\end{equation}
where for $u \subseteq \{1,\dots,d\}$ we denote by $|u|$ its cardinality, 
and $(\vz_{u},\mathbf{1})$ denotes the vector 
containing the coordinates of $\vz$ with indices in $u$, and the other 
coordinates set equal to $1$. 

In this case the reproducing kernel is given by 
\[
K_{d,\gamma}(\vz,\vy)=\prod_{j=1}^{d}(1+\gamma_{j}
[1-\max(z_{j},y_{j})])\quad \text{ for } \vz,\vy \in[0,1]^{d}.
\]
The weighted tensor product Sobolev space allow for explicit QMC constructions 
deriving error estimates of the form
\begin{equation}\label{rate}
e_{N}(\mathbb{F}_{d})\leq C(\delta)N^{-1+\delta}
\quad \text{ for } \delta\in(0,\textstyle{\frac{1}{2}}],
\end{equation}
where the constant $C(\delta)$ is independent on the dimension $d$,
if the sequence of weights $(\gamma_{j})$ satisfies the condition (see \cite{Kuo2003})
\[
\sum_{j=1}^{\infty}\gamma_{j}^{\frac{1}{2(1-\delta)}}<\infty\,.
\]
Traditional unweighted function spaces considered for integration 
suffer from the curse of dimensionality. Their weighted variants describe a 
setting where the variables or group of variables may vary in importance, corresponding 
to an anisotropic problem. Many integration problems in 
practice start with an isotropic setting but can 
be modified to an anisotropic one using a proper transformation. 
The concentration of importance in few variables or groups of few variables 
gives a partial explanation of why some 
very high-dimensional spaces become tractable for QMC.\\
Explicit QMC constructions satisfying \eqref{rate} are for example \textit{shifted lattice rules} 
for weighted spaces \cite{Kuo2003}.  
The rate (\ref{rate}) can be also obtained for Niederreiter and Sobol' sequences (see \cite{Wang03}).
The idea of ``weighting'' the norm of the spaces to obtain tractable results can be applied in fact to  
more general function spaces than smooth function spaces of tensor product form, and many integration 
examples can be found in \cite{Novak_and_Wozniakowski2}. 
In our numerical experiments, we used so far QMC algorithms based on 
a particular type of low-discrepancy sequences. 
Numerical experiments with shifted lattice rules will be carried out in the near future, following 
new techniques for fixing adequate weights introduced in \cite{GLLZ12}.

\section{Low-discrepancy \texorpdfstring{$(t,d)$}{(t,d)}-sequences}\label{t_d_seq}

The most well known type of low-discrepancy sequences are the so-called
$(t,d)$-sequences. 
To introduce how $(t,m,d)$-nets and $(t,d)$-sequences are defined, 
we consider first \textit{elementary intervals} in a integer base $b \ge 2$.
Let $E$ be any sub-interval of $[0,1)^{d}$ of the form 
$E=\prod_{i=1}^{d}[a_{i}b^{-c_{i}},(a_{i}+1)b^{-c_{i}})$
with  $a_{i},\: c_{i} \: \in \mathbb{N} ,  c_{i} \ge 0, \:0\leq a_{i} < b^{-c_{i}}$ 
for $1\leq i\leq d$. An interval of this form is 
called an elementary interval in base $b$.
\begin{definition}
 Let $\:0\leq t \leq m$ be integers. A $(t,m,d)$-net in base $b$ is a point
 set $P_N$ of $N=b^{m}$ points in  $[0,1)^{d}$ 
such that  every elementary interval 
$E$ in base $b$ with $\lambda_{d}(E)=\frac{b^{t}}{b^{m}}$ 
contains exactly $b^{t}$ points.
\end{definition}

\begin{definition}
 Let $t\geq 0$ be an integer. A sequence $\vz_{1},\vz_{2},...$ 
of points in  $[0,1)^{d}$ is a $(t,d)$-sequence in base $b$ if for all integers 
$k\geq0$ and $m>t$, the point set consisting of $N=b^{m}$ points 
$\vz_{i}$ with $kb^{m}\leq i < (k+1)b^{m}$,   
is a $(t,m,d)$-net in base $b$.
\end{definition}
The parameter $t$ is called the \textit{quality parameter} 
of the $(t,d)$--sequences.
In \cite{Nied92}, theorem 4.17, it is shown that $(t,d)$-sequences 
are in fact low-discrepancy sequences. We reproduce this result in the following
\begin{theorem}
The star-discrepancy $D^{\star}$ of the first $N$ terms $P_N$ of a $(t,d)$-sequence in 
base $b$, satisfies
\[
 N D^{\star}(P_N) \leq C(d,b) b^t (log(N))^d + O(b^t (log(N))^{d-1}),
\]
where the implied constants depend only on $b$ and $d$.
If either $d=2$ or $b=2$, $d=3,4$, we have
\[
C(d,b)=\frac{1}{d}\left( \frac{b-1}{2 log(b)} \right)^d,
\]
and otherwise
\[
C(d,b)=\frac{1}{d!} \frac{b-1}{2 \lfloor b/2 \rfloor} \left( \frac{\lfloor b/2 \rfloor}{log(b)} \right)^d. 
\]
\end{theorem}

Explicit constructions of $(t,d)$-sequences are available. Examples are 
the generalized Faure, Sobol', Niederreiter and Niederreiter--Xing sequences. 
All these examples fall into the category of constructions 
called \textit{digital sequences}, see \cite{DiPi10}. 
To complete this section, we will describe briefly Sobol' sequences and give references for their practical 
implementations. Sobol' sequences are among the most widely used and recommended QMC sequences 
by simulation practitioners (see \cite{Glasserman} for successful applications of Sobol' sequences in finance),
and they are the QMC sequences selected for our numerical experiments.

\subsection{Sobol' sequences and implementations}\label{Sobol_impl}
The pioneering work of Sobol' \cite{SOB67} introduced the first known construction of $(t,d)$-sequences, 
and they can be viewed now as a special case of the so-called generalized Niederreiter sequences 
in base $b=2$ (see chapter 8 in \cite{DiPi10} and references therein).
The basic construction of Sobol' sequences can be described as follows:
\begin{enumerate}
 \item Let $p_1,\dots,p_d$ be primitive polynomials of degree $deg(p_i)=:e_i$ over the field $\mathbb{F}_2[x]$, 
\[
p_i(x)=x^{e_i} + a_{1,i}x^{e_i -1} +a_{2,i}x^{e_i -2}+\cdots+a_{e_i-1}x +1 \quad \text{ for } 1\le i \le d,
\]
sorted according to their degree in increasing order. 
\item Let $1\le m_{1,i},\dots,m_{e_i,i}$ be odd natural numbers with $m_{k,i}<2^k$ for $1\le k\le e_i$, $1\le i \le d$. Then define for 
$k>e_i$ recursively
\[
 m_{k,i}=2a_{1,i}m_{k-1,i} \oplus \cdots \oplus 2^{e_i-1}a_{e_i-1}m_{k-e_i+1,i}\oplus 2^{e_i}m_{k-e_i,i}\oplus m_{k-e_i,i} 
\quad \text{ for } 1\le i \le d,
\]
where the operator $\oplus$ is the bit-by-bit exclusive-or operator.
\item Define the \textit{direction numbers} $v_{k,i}$ by 
\[
v_{k,i}= \frac{m_{k,i}}{2^k} 
\quad \text{ for } k\ge 1, \:1\le i \le d. 
\]
\item Consider a natural number $n$ with binary expansion $n=n_0 + n_1 2 + \cdots + n_{r-1} 2^{r-1}$, and define 
\[
z_{n,i}=n_0v_{1,i} \oplus n_1v_{2,i} \oplus \cdots \oplus n_{r-1}v_{r,i} 
\quad \text{ for } 1\le i \le d. 
\]
\item Finally consider the sequence of points $\vz_0,\vz_1,\dots$ in $[0,1)^d$ defined by
\[
\vz_n=(z_{n,1},\dots,z_{n,d})\;.
\]

\end{enumerate}
A sequence of points $\vz_0,\vz_1,\dots$ in $[0,1)^d$ as defined above is called a Sobol' sequence. 
The quality parameter of Sobol' sequences is given by 
\[
t=\sum_{i=1}^{d}(e_i-1). 
\]
The direction numbers $v$ defined above determine the quality of low dimensional projections of the points 
in the Sobol' sequences. Sobol' \cite{SOB67} introduced an additional uniformity condition called \textit{Property A} 
in order to give a criteria for selection of initial $m$ numbers in the recurrence stated above. Efficient implementation of 
Sobol' sequences are based on \textit{Gray code}.
A classical reference for practical implementation  
is \cite{BratleyFox88}. Joe and Kuo \cite{KuoJoe1}, \cite{JoeKuo2}  give an alternative for 
selection of direction numbers based in a \textit{weighted} approach, focused in a setting where the importance of variables decay as their 
dimension number increase. Moreover, one can find a three-pages-note with a 
short and simple description on implementation of Sobol' sequences at 
the website \url{http://web.maths.unsw.edu.au/~fkuo/sobol/index.html}. 
New developments of Sobol' sequences and comparison between available implementations can be found in \cite{WILM:WILM10056}.

\section{Randomized QMC }\label{sec:RQMC}

There are some advantages in retaining the probabilistic properties of the sampling.  
There are practical hybrid methods permitting us to combine the good features of MC and 
QMC. Randomization is an important tool for QMC if we are interested for a practical
error estimate of our sample quadrature $Q_N$ to the desired integral. One goal is 
to randomize the deterministic point set $P_N$ generated by QMC in a way that 
the estimator $\hat{Q}_N$ preserves unbiasedness. Another important goal is to preserve 
the better equidistribution properties of the deterministic construction. 
 
The simplest form of randomization applied to \textit{digital sequences} seems to be 
the technique called \textit{digital $b$--ary shifting}. In this case, we add 
a random shift $\Delta \in [0,1)^d$ to each point of the deterministic set 
$P_N=\{\vz_{1},...,\vz_{N}\}$ using 
operations over the selected ring $\mathbb{F}_b$. 
The application of this randomization preserves in particular the $t$ value of any projection of 
the point set (see \cite{L'Ecuyer01} and references therein). The resulting estimator is 
unbiased.\\
The second randomization method we consider is the one introduced by  
Owen (\cite{OWE95}) in 1995. He considered $(t,m,d)$-nets and $(t,d)$-sequences 
in base $b$ and applied a randomization procedure based on permutations of the digits of 
the values of the coordinates of points in these nets and sequences. This can be interpreted 
as a random scrambling of the points of the given sequence in such a way 
that the net structure remains unaffected.
We do not discuss here in detail Owen's randomization procedure, 
or from now on called \textit{Owen's scrambling}. 
The main results of this randomization procedure can be stated in the following
\begin{proposition}(\textbf{Equidistribution})\\
A randomized $(t,m,d)$-net in base $b$ using Owen's scrambling is again a $(t,m,d)$-net 
in base $b$ with probability 1. A randomized $(t,d)$-sequence in base $b$ using Owen's 
scrambling is again a $(t,d)$-sequence in base $b$ with probability 1.
\end{proposition}
\begin{proposition}(\textbf{Uniformity})\\
Let $\tilde{\vz}_i$ be the randomized version of a point 
$\vz_i$ originally belonging  to a  $(t,m,d)$-net 
in base $b$ or a $(t,d)$-sequence in base $b$, using Owen's scrambling. 
Then $\tilde{\vz}_i$ has 
the uniform distribution in $[0,1)^d$, that is, for any Lebesgue measurable set $G \subseteq 
[0,1)^d$ , $P( \tilde{\vz}_i \in G)= \lambda_d(G)$, 
with $\lambda_d$ the $d$-dimensional Lebesgue measure.  
\end{proposition} 

The last two propositions state that after \textit{Owen's scrambling} of \textit{digital sequences} 
we retain unaffected the low-discrepancy properties of the constructions, and that 
after this randomization procedure we obtain random samples uniformly distributed in $[0,1)^s$. \\

The basic results about the variance of the randomized QMC estimator $\hat{Q}_N$ 
after applying \textit{Owen's scrambling} 
to $(t,m,d)$-nets in base $b$ (or of $(t,d)$-sequences in base $b$ ) 
can be found in \cite{Owen97}. We summarize these results in the following

\begin{theorem}
Let $\tilde{\vz}_i$, $1\le i \le N$, be the points of a 
scrambled $(t,m,d)$-net in base $b$, and let $f$ be a function 
on $[0,1)^d$ with integral $I$ and variance $\sigma^2=\int (f-I)^2 d\vz  < \infty.$ 
Let $\hat{Q}_N= N^{-1} 
\sum_{i=1}^N f(\tilde{\vz}_i)$, where $N=b^m$. 
Then for the variance $V(\hat{Q}_N)$ of the randomized QMC estimator 
it holds
\[ V(\hat{Q}_N)=o(N^{-1}), \: \text{ as } N \rightarrow \infty,  \quad \text{and} \quad 
 V(\hat{Q}_N)\leq \frac{b^t}{N}\left( \frac{b+1}{b-1} \right)^d \sigma^2.\]
For $t=0$ we have
\[ V(\hat{Q}_N)\leq \frac{1}{N}\left( \frac{b}{b-1} \right)^{d-1} \sigma^2.\]
\end{theorem}
The above theorem says that the variance of the randomized QMC estimator $\hat{Q}_N$ 
using scrambled $(0,m,d)$--nets is always smaller than 
a small multiple of the variance of the corresponding MC estimator.
If the integrand at hand is smooth enough, using \textit{Owen's scrambling} 
it can be shown that one can obtain an improved asymptotic error estimate of order
$O(N^{-\frac{3}{2}-\frac{1}{d}+\delta})$, for any $\delta>0 $, see \cite{Owen08}. 
Improved scrambling techniques have been developed in \cite{MAT98},\cite{Tezuka03}.\\

\section{Effective dimensions and sensitivity indices}\label{eff_sen}

In many practical applications, one encounters functions for which 
the total variance is concentrated
in a small part of its ANOVA terms. The notion of \textit{effective 
dimension} of a function 
was first introduced in \cite{CAF97} to describe the contribution 
of a group of variables to the 
total variance. 

\subsection{ANOVA Decomposition}
Using the \emph{ANOVA (Analysis of Variance) decomposition} we decompose a function into 
a sum of simpler functions, see \cite{Sobol2001}. 
Let $D=\{ 1,\dots,d\}$. 
For any subset $\mathbf{i} \subseteq D$, let $|\mathbf{i}|$ denote its cardinality and 
$(D-\mathbf{i})$ be its complementary set in $D$. 
Let $\vz_\mathbf{i}=(z_j:\;j\in\mathbf{i})$ be the $|\mathbf{i}|-$dimensional vector 
containing the coordinates of $\vz$ with indices in $\mathbf{i}$. 
Now assume that $f$ is a square integrable function. 
Then we can write $f$ as the sum of $2^d$ ANOVA terms:
\[
f(\vz)=\: \sum_{\mathbf{i}\subseteq D}f^\mathbf{i}(\vz)\;,
\]
where the ANOVA terms $f^\mathbf{i}(\mathbf{x})$ are defined recursively by
\[
f^\mathbf{i}(\vz)= \:
\int_{[0,1]^{d-|\mathbf{i}|}}f(\vz_{\mathbf{i}},\vz_{D-\mathbf{i}})
d\vz_{D-\mathbf{i}}
	\:-\:
	\sum_{\mathbf{j}\subsetneq\mathbf{i}}f^\mathbf{j}(\vz)\;,
\]
and $f^\varnothing = I(f)$. The sum of the right--hand side is over strict subsets
$\mathbf{j}\neq \mathbf{i}$, and we use the convention
$\int_{[0,1]^{0}}f(\vz)d\vz_\varnothing =f(\vz)$.
The ANOVA terms enjoy the following interesting properties:
\begin{enumerate}
\item $\int_0^1 f^\mathbf{i}(\vz)dz_j=0$ for $j\in\mathbf{i}$.
\item The decomposition is orthogonal, in that
	$\int_{[0,1]^d}f^\mathbf{i}(\vz)f^\mathbf{j}(\vz)d\vz=0$ 
whenever $\mathbf{i}\neq\mathbf{j}$.
\item Let $\sigma^2(f)= \int_{[0,1]^{d}}f(\vz)^2\,d\vz \:-\:(I(f))^2$ 
	be the variance of $f$, then we have: 
	\[
	\sigma^{2}(f) 
	\: = \: 
	\sum_{\mathbf{i}\subseteq D}\sigma^{2}_{\mathbf{i}}(f),
	\quad\text{ where }\quad
	\sigma^2_{\mathbf{i}}(f)=\int_{[0,1]^{d}}f^\mathbf{i}(\vz)^2\,d\vz
	\] 
	for $|\mathbf{i}|>0$ is the variance of $f^\mathbf{i}$ and 
	$\sigma^2_{\varnothing}(f)=0$.
\end{enumerate}

\begin{definition}~
\begin{enumerate}
\item $f$ is said to have effective dimension in the superposition 
sense $d_s$ with proportion $p$, for 
$0<p<1$, if $d_s$ is the smallest integer that satisfies
\[ \;\sum_{|\mathbf{i}|\leq d_s } \sigma^{2}_{\mathbf{i}}(f) \ge p \sigma^{2}(f). \]
\item $f$ is said to have effective dimension in the truncation 
sense $d_t$ with proportion $p$, for 
$0<p<1$, if $d_t$ is the smallest integer that satisfies
\[ \;\sum_{\mathbf{i} \subseteq \{1,\dots,d_t\} } \sigma^{2}_{\mathbf{i}}(f) \ge p \sigma^{2}(f). \]
\end{enumerate}
\end{definition}

One can estimate the effective dimension in truncation sense based on the 
algorithm proposed by Wang and Fang \cite{WaFa03}. They show that the following equality holds
\[
\int_{[0,1]^{2d-|u|}}f(\vz)f(\vz_{u},\mathbf{y}_{D-u})
d\vz d\mathbf{y}_{D-u} =
\sum_{\mathbf{i}\subseteq u}\sigma^{2}_{\mathbf{i}}(f) + f^\varnothing. 
\]
Thus, for estimating the effective dimension in truncation sense, we need to estimate the following 
tree type of integrals
\begin{equation}\label{sec:Eff:Eff}
\int_{[0,1]^{d}}f(\vz)d\vz,  \quad \int_{[0,1]^{d}}f^2(\vz)d\vz, \quad 
\int_{[0,1]^{2d-|u|}}f(\vz)f(\vz_{u},\mathbf{y}_{D-u})
d\vz d\mathbf{y}_{D-u},
\end{equation}
for $u=\{1,\dots,l\}, \; l=1,2,\dots$, using MC or QMC, until the 
proportion of variance defining the effective dimension is reached. 
In many applications, the proportion value is usually taken as $p=0.99$.

Given any nonempty family $T$ of subsets of $D$, we can consider the function defined 
by the corresponding ANOVA terms $f_{T}(\vz)\;:=\;\sum_{\bi \in T } f^{\bi}(\vz)$. 
For example, given a fixed proportion value $p$ we can consider the sets $T=\{\mathbf{i}: \mathbf{i}\subseteq\{1,\dots,d_t\} \}$ 
or $T=\{\mathbf{i}:|\mathbf{i}|\leq d_s\}$  to define the \textit{effective part} $f_{T}$ of the function $f$ in truncation or 
superposition sense respectively.
The integration error for $f$ of a QMC algorithm $Q_{N}$ can be bounded then by 
\begin{equation}\label{errorsplit}
 |I(f)- Q_{N}(f)|
\leq 
|I(f_{T})- Q_{N}(f_{T})| 
  + |I(f-f_{T})- Q_{N}(f-f_{T})|.
\end{equation}
If $f_{T}$ is the effective part of a function exhibiting low-effective dimension in superposition or truncation sense, 
then the second error term in the right hand side of \eqref{errorsplit} represents the integration error over the \textit{rest function} $f-f_{T}$ having a relatively small 
variance. For many practical applications, the second error term in \eqref{errorsplit} 
is believed to be so small that can be neglected (see \cite{Wang_Sloan_05}).

If the truncation effective dimension is small, then few variables are important for 
sampling. If the superposition effective dimension is small, say $d_s$ equals $2,3 \text{ or maybe } 4$, 
then some QMC sequences and their randomizations 
are also expected to outperform MC, because they can exhibit much better equidistributed 
low-dimensional projections than MC (see \cite{Wang_Sloan_05},\cite{WILM:WILM10056}).\\
The ordering of the variables of the integrand is important for achieving a reduction of the 
effective dimension in the truncation sense $d_t$, and usually affects the performance of 
QMC and their randomizations in practice. Sensitivity indices usually help to 
order the variables in a convenient way for integration with QMC.
   
\subsection{Derivative based sensitivities}\label{sec:EffDim:sensi}
As pointed out by Sobol' and Kucherenko in \cite{SobK09}, 
very often derivative based measures of sensitivities can successfully be used for 
detecting non essential variables. Small values of first order derivatives 
of a function implies small values of one--dimensional total Sobol' sensitivity indices. 
Let $\sigma^2_{\bi}(f)$ denote the partial variance corresponding to the ANOVA term $f^{\bi}$. 
Define
\[\sigma^2_{\{j\}}(f)^{\text{tot}}= \sum_{\bi \subset D : j \in \bi} \sigma^2_{\bi}(f),\]
then it is shown in \cite{Sobol2001} and \cite{SobK09} that
\[\displaystyle \sigma^2_{\{j\}}(f)^{\text{tot}}=\frac{1}{2} \int_{[0,1]^d} 
\int_0^1 [f(\vz)-f(z_1,\cdots,z_{j-1},z_j',z_{j+1},\cdots,z_n)]^2 d\vz dz_j',\]
from which one obtain the following two results:
\begin{enumerate}
\item if $c<|\frac{\partial f}{\partial z_j}|<C$, then 
\[\frac{c^2}{12}\le \sigma^2_{\{j\}}(f)^{\text{tot}}\le \frac{C^2}{12}\;, \]
\item and if $\frac{\partial f}{\partial z_j} \in L_2([0,1]^d)$, then
\begin{equation}\label{variance_diffbound}
\sigma^2_{\{j\}}(f)^{\text{tot}}\le\frac{1}{\pi^2} \int_{[0,1]^d} 
\left(\frac{\partial f}{\partial z_j}(\vx)\right)^2 d\vz. 
\end{equation}
\end{enumerate}
As a consequence of the bounds stated above, the total variance corresponding to non--essential 
variables of a function can be bounded using first order derivatives information. 
In a wide variety of problems in practice, the gradient of a scalar function can be efficiently computed 
through algorithmic differentiation (see \cite{AD}), at a cost at most 4 times of 
that for evaluating the original function. Thus, a cheap method for estimating
 derivative based sensitivities, and an upper bound on the effective
dimension in the truncation sense (as stated in the following simple
Proposition), may be available using 
algorithmic differentiation. The variance is, clearly, invariant to a permutation of the 
variables. This allowed us to consider the following
\begin{definition}
Given an bijection (permutation) $\pi:\{1,\dots,d\} \rightarrow \{1,\dots,d\}$, 
$f$ is said to have $\pi$--effective dimension in the truncation 
sense $d_t$ with proportion $p$, for 
$0<p<1$, if $d_t$ is the smallest integer that satisfies
\[ \;\sum_{\bi \subset \{\pi^{-1}(1),\dots,\pi^{-1}(d_t)\} } \sigma^{2}_{\bi}(f) \ge p \sigma^{2}(f). \]
\end{definition}
\begin{proposition}
Let $f \in L_2([0,1]^d)$ such that $\frac{\partial f}{\partial z_j} 
\in L_2([0,1]^d)$ $\forall$ $1\le j \le n$. Consider the derivative based 
sensitivities \[v_i:=\frac{1}{\pi^2} \int_{[0,1]^d} 
\left(\frac{\partial f}{\partial z_j}(\vz)\right)^2 d\vz,\]
and consider any permutation $\pi^*:\{1,\dots,d\} \rightarrow \{1,\dots,d\}$ such that 
\[v_{(\pi^*)^{-1}(k)}\ge v_{(\pi^*)^{-1}(k+1)} , \quad \forall \,1 \le  k \le d-1,\]
(a non-increasing ordering of the sensitivities $v_i$'s, resulting in what is 
called by the authors a "Diff--decay--ordering" $\pi^{*}$).\\
Let $0<p<1$ be a fixed proportion parameter. If there exists an integer $m$ such that
\begin{equation}\label{sec:Eff:Prop}
 \sum_{j=m+1}^{d} v_{(\pi^*)^{-1}(j)} \le (1-p) \sigma^{2}(f) 
\end{equation}
then, it follows that the $\pi^*$--effective dimension in the truncation 
sense with proportion $p$ is at most $m$. 
\end{proposition}
\begin{prf}~
 
Let $d_t$ denote the $\pi^*$ --effective dimension in the truncation 
sense with proportion $p$. Consider $m$ satisfying \eqref{sec:Eff:Prop} 
and define $T_m=\{\bi: \bi \subset \{(\pi^*)^{-1}(1),...,(\pi^*)^{-1}(m)\}\}$ 
and $ f_{T_m}=\, \sum_{\bi \in T_m} f^{\bi}(\vz)$. 
It follows from the $L_2$ orthogonality of ANOVA decomposition and 
\eqref{variance_diffbound} that 
\begin{align*}
&\sigma^2(f)-\sigma^2(f_{T_m})=\sigma^2(f-f_{T_m})=
\sum_{\{\bi:\bi \subset D \wedge \bi \not \in T_m  \}} \sigma_{\bi}^2(f)
\le \sum_{j=m+1}^d \sum_{\{\bi \subset D : (\pi^*)^{-1}(j) \in \bi\}} \sigma_{\bi}^2(f)\\
&= \sum_{j=m+1}^d \sigma^2_{\{(\pi^*)^{-1}(j)\}}(f)^{\text{tot}} \le 
\sum_{j=m+1}^{d} v_{(\pi^*)^{-1}(j)} \le (1-p) \sigma^{2}(f).
\end{align*}
It follows $\sigma^2(f_{T_m})\ge p \sigma^{2}(f)$ and thus $d_t \le m$, what was required to be proved.

\end{prf}

\section{Weighted uniform sampling}\label{sec:WUS}

In this section we will discuss the method we used to approximate observables as they are defined in equation \eqref{sec:Plain:OBS}.
Before the WUS method can be applied to this expression, it is necessary to perform a transformation of the variables $x_i$ to the $d$-dimensional unit cube, $[0,1]^d$. In the cases we will consider in section \ref{sec:Num_Exp}, this transformation will always be of the form
\begin{equation}
\label{eq:gen_gaussian_trafo}
x_i = \sum_j A_{ij} \Phi^{-1}(z_j) \; ,
\end{equation}
with $A$ being a positive definite matrix and $\Phi^{-1}$ the inverse of the PDF of the standard normal distribution.
After the transformation equation \eqref{sec:Plain:OBS} reads:
\begin{gather}
\label{eq:trafo_to_unit_cube}
\langle O \rangle = \frac{
\int_{[0,1]^d} O( A \Phi^{-1}(\vz) ) W(\vz) dz_1 \ldots dz_d
}{
\int_{[0,1]^d}  W(\vz)  dz_1 \ldots dz_d
}\\
\nonumber
W(\vz) = \exp \left[ -  S( A \Phi^{-1}(\vz) ) + \frac{1}{2}\sum_{i} ( \Phi^{-1}(z_i) )^2 \right] \; .
\end{gather}
Now, in the WUS method   
points $\vz_j, \; 1\le j \le N$, 
are generated from a uniform distribution in $[0,1]^d$. 
Using these points, a quotient of integrals of the form 
\[
Q(f_1,f_2) := R :=\frac{ \int_{[0,1]^d}f_1(\vz)d\vz}{ \int_{[0,1]^d}f_2(\vz)d\vz}
\] 
can then be approximated by taking the rule 
\begin{equation}
\label{eq:WUS}
Q_N(f_1,f_2):=\frac{ \sum_{j=1}^N f_1(\vz_j)}{\sum_{j=1}^N f_2(\vz_j)} \; ,
\end{equation}
where the functions $f_i$ could be of very general, in particular non-Gaussian nature.
For our example these functions can be read off from equation \eqref{eq:trafo_to_unit_cube}: $f_1 = O( A \Phi^{-1} (\vz) ) W(\vz)$ and $f_2 = W(\vz) $.
For the case that $W(\vz)$ is really a function of $\vz$ (and not just a constant), this way of evaluating integrals over certain weight functions $W$ 
is known as reweighting technique in field theory or statistical 
physics. 
A crucial element of the WUS (reweighting) method is that the sampling 
points have a large enough overlap with the weight functions 
$f_i$ considered. 
The resulting WUS estimator $\hat{Q}_N(f_1,f_2)$ from \ref{eq:WUS} has been analyzed in \cite{PowellSwann66} and applications have been 
investigated for example in \cite{SpaMa} and \cite{Caflisch95}. The bias and the root mean 
square error (RMSE) of this estimator satisfy 
\begin{align*}
& Bias(\hat{Q}_N(f_1,f_2))=\frac{R \,var(f_2)}{N} -\frac{cov(f_1,f_2)}{N} + O(N^{-\frac{3}{2}}) \\
& RMSE(\hat{Q}_N(f_1,f_2))=\frac{\sqrt{var(f_1) + R^2 var(f_2) - 2R\, cov(f_1,f_2)}}
{\sqrt{N}} + O(N^{-\frac{3}{4}}) \; .
\end{align*}
The bias of the estimator in this case is asymptotically negligible compared with the RMSE.

A deterministic version of the WUS estimator has been considered in \cite{SpaMa}. 
In particular, it follows from Theorem 4.2 in \cite{SpaMa} that if the integrands $f_1,f_2$ 
in \eqref{eq:trafo_to_unit_cube} are of bounded 
variation in the sense of Hardy and Krause, then by the use of a low-discrepancy sequence $\vz_{1},\vz_{2},...$ 
instead of i.i.d. uniform random samples we obtain the integration error asymptotic  
\[\left| Q_N(f_1,f_2) -Q(f_1,f_2) \right|=     
O(N^{-1}(\log(N))^{d}).
\] 
Similar results for the bias and RMSE of $\hat{Q}_N(f_1,f_2)$ considering randomized QMC sequences 
instead of i.i.d. random samples are not known to the authors. 
Nevertheless, the numerical results in section \ref{sec:Num_Exp} (e.g. estimated ground state energy vs. theoretical values given in \cite{Blank79})
seem to indicate that in our examples the bias under scrambled Sobol' sequences is very small and has no practical relevance.

One clear disadvantage of WUS against Mc-MC or Importance Sampling 
for problems with large regions of relative low values 
of the integrands is that with WUS we sample over 
the entire unit cube $[0,1]^d$ uniformly,
thus the method is dependent on how we transformed the problem to the unit
cube. In contrast, Mc-MC Importance Sampling based techniques for
models in high-energy or statistical physics usually focus on characteristic
or important regions of the integrands aiming to sample directly from the
underlying distribution of the problem, using in this way 
only the most relevant sample points.  

\section{Numerical experiments}
\label{sec:Num_Exp}
We consider for our numerical tests the \textit{quantum mechanical 
harmonic and anharmonic oscillator} in the \textit{path integral
approach}  as described in section 2.
For definiteness we repeat here the expression for the action of the system:
 \begin{equation}
\label{eq:action_detail}
 S(\vx)=\frac{a}{2} \sum_{i=1}^d \left( \frac{M_0}{a^2} (x_{i+1}-x_i)^2 + \mu^2 x_i^2 
 + 2 \lambda x_i^4 \right) \; .
 \end{equation}
We investigate the two observable functions
\[
O_1(\vx)=\frac{1}{d}\sum_{i=1}^d x_i^2 \, , \;
O_2(\vx)=\frac{1}{d}\sum_{i=1}^d x_i^4 \; ,
\]
using the notation $\left\langle X^2 \right\rangle$,$\left\langle X^4 \right\rangle$ 
for $\left\langle O_1(\vx) \right\rangle$,$\left\langle O_2(\vx) \right\rangle$ in our tests. 
In addition, we will look at the ground state energy $E_0$ which, by virtue of 
the virial theorem, is related to $O_1$ and $O_2$ by $E_0 =  \mu^2  O_1 + 3 \lambda  O_2 + \frac{\mu^4}{16}$.\\
Furthermore, we provide a programme for the QMC simulation of the (an)harmonic oscillator at \url{http://arxiv.org/format/1302.6419v3} (``Source'') and give a description of the usage of the programme in the appendix (section 10).

\subsection{Harmonic Oscillator}
\label{ssec:HO}
\begin{figure}[ht]
\centering
  \begin{minipage}[b]{0.8\linewidth}
    \centering
  \includegraphics[width=\textwidth]{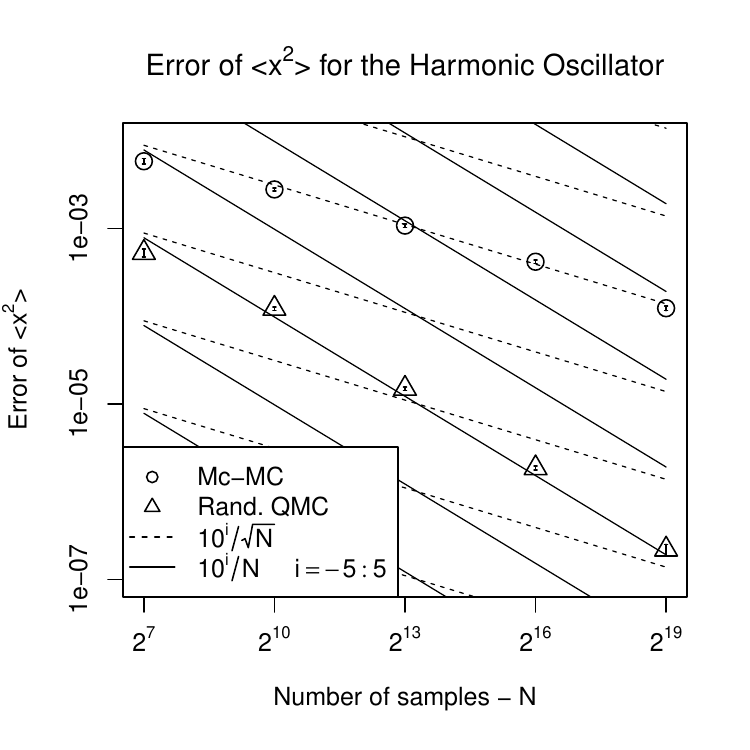}
  \caption{The error of $\langle X^2 \rangle$ in dependence on the number of 
samples $N$. The parameters here were chosen as $\lambda=0$ (harmonic oscillator),
$d=51$, $M_0=0.5$ and $\mu^2=2.0$. The error of the error was obtained by repeating 
the numerical experiment 30 times, see also the text.}
  \label{fig:x2_harmonic}
\end{minipage}
\end{figure}

For the harmonic oscillator we can apply immediately the direct sampling approach 
described in sections \ref{sec:Plain:Int} and \ref{sec:WUS} for calculating 
estimates of observables $O(.)$ by setting
\[
f_1 = O( A \Phi^{-1}(\vz) ) \; , \;\; f_2 = 1
\]
in \eqref{eq:WUS}.
The matrix $A$ is a square root of $C$, the covariance matrix of the variables $x_i$, 
appearing in the action if it is expressed as a bi-linear form: $S(\vx)=\frac{1}{2}\vx^T C^{-1} \vx$, written explicitly
\begin{equation}
C^{-1}_{ij} = \frac{2 M_0}{a} \left[ u \delta_{ij} - \frac{1}{2} \left( \delta_{i+1\;j} + \delta_{ i \; j+ 1 } \right) \right] ,\; u = 1+\frac{a^2\mu^2}{2 M_0} \; .
\label{eq:cij}
\end{equation}
Different factorizations, namely Cholesky and PCA (principle component analysis) have been tried out. 
The PCA based factorization turned out to perform better in our tests, which is the reason why we will only show results for this method.
Note that, independently of which factorization we have chosen for $C$, for the case of the harmonic oscillator we sample 
directly from a Gaussian distribution and the 
considered observables functions $O(.)$ are just multivariate polynomials of low
degree. Thus, the effective dimension in the superposition sense of the resulting non-constant integrand $f_1=O( A \Phi^{-1}(\vz))$
is upper bounded by the highest degree of the polynomials defining the observables $O(.)$ (this is true because the 
ANOVA decomposition is known to retain a \textit{minimal} representation \cite{KuoSWW10}). 
Therefore the problem has intrinsic low-effective dimension in the superposition sense, 
and is expected that (randomized) QMC outperforms MC in this case. The PCA 
factorization seems to achieve further improvements since it can reduce, 
in addition, the effective dimension in the truncation sense. This is usually the case for 
Gaussian integrands considered in mathematical finance, involving a covariance matrix with rapid decaying 
eigenvalues (see \cite{Glasserman},\cite{Wang_Sloan_05}).   
The PCA factorization can be explicitly obtained for circulant Toeplitz matrices 
and the matrix--vector products can be efficiently computed by means of the fast Fourier transform.
Given that the covariance matrix $C$ is circulant Toeplitz, we have that 
$C=G\Lambda G^T$, with $G:=Re(F)+Im(F)$,
\begin{equation}
  (F)_{kl} = \frac{1}{\sqrt{d}}e^{-\frac{2\pi i}{d} k l} 
\end{equation}
being the Fourier matrix and $\Lambda$ the diagonal 
matrix of positive eigenvalues (Lemma 4 in \cite{Graham20113668}). 
Thus $A=G\Lambda^{\frac{1}{2}}$ is a factorization of $C$, 
and in this case one can follow a recipe for generating normals with randomized QMC based on 
the discrete Fourier transform and using 
fast Fourier transform (FFT) techniques as described in \cite{Graham20113668}:
\begin{enumerate}
\item Generate a randomized QMC point $\tilde{\vz}$.
\item Compute $\tilde{\mathbf{y}}=\Phi^{-1}(\tilde{\vz})$.
\item Compute $\tilde{\mathbf{w}}=(\sqrt{\coveval_1}\tilde{y}_{\pi^{-1}(1)}, 
      \dots,\sqrt{\coveval_d}\tilde{y}_{\pi^{-1}(d)})$, where \\
\begin{gather}
\coveval_j = \left[ \frac{ 2 M_0 } {a} \left(  u - \cos ( 2\pi j/d ) \right) \right] ^{-1}, \; 1\le j \le d
\end{gather}
are the eigenvalues 
      in the diagonal matrix $\Lambda$, and $\pi(.)$ is a fixed permutation of the variables.
\item Compute $\tilde{ \mathbf{v}}=\text{FFT}(\tilde{\mathbf{w}})$.
\item Take $\tilde{\mathbf{x}}=Re(\tilde{ \mathbf{v}})+Im(\tilde{ \mathbf{v}})$ as the resulting 
point sample.
\end{enumerate}
It is (strongly) recommended to fix first the permutation $\pi(.)$ 
such that $(\coveval_{\pi(j)})_{j=1}^d$ are 
in non-increasing order, and this permutation was taken in our experiments. 
If this permutation of variables does not lead to satisfactory results, 
the analysis described in \ref{sec:EffDim:sensi} can be carried out to investigate if 
a possible different permutation leads to more effective dimension reduction and better results.\\
In the ordinary Mc-MC approximation, we used the Mersenne Twister\cite{Matsumoto98} pseudo random number generator.
We note in passing that the Mc-MC samples were generated in exactly the same way 
as described above for randomized QMC with the only difference that in step
one Mc-MC points were generated according to the Gaussian measure 
of the harmonic oscillator. 
This corresponds to a heatbath algorithm, where all variables $x_i$ are updated at the same time and a reweighting procedure in the anharmonic case.
For the QMC tests, we use the Sobol' sequences described in section \ref{Sobol_impl} from \cite{KuoJoe1}, 
with the random scrambling technique proposed by J. Matous\v{e}k\cite{MAT98}.
The error of $\langle X^2 \rangle$ was obtained by scrambling 10 times the QMC sequence and making 10 runs of an Mc-MC simulation (with different seeds). 
This procedure is repeated 30 times in both cases to obtain the error of the error.
From the results shown in figure \ref{fig:x2_harmonic}, we can see a scaling of 
the errors $N^{-1/2}$ for Mc-MC and $N^{-1}$ for randomized QMC, for large $N$.
Although this example is trivial, it was our first successful application of the QMC approach in a 
physical lattice model and motivated us to pass on to more complicated models.

\subsection{Anharmonic Oscillator}

The WUS (reweighting) approach was also used for the problem of the 
anharmonic oscillator to estimate 
$\langle X^4 \rangle$, $\langle X^2 \rangle$ and the 
ground state energy of the system ($ E_0 $).
With the anharmonic term in action, the probability distribution 
function (PDF) of the variables $x_i$  
is of non-Gaussian nature and hence 
becomes very complicated.       
This makes it very hard to generate the samples directly from the 
PDF of the anharmonic oscillator. 
Instead of this, we consider the WUS method with samples originated from an  
importance density (see (4.6) in \cite{SpaMa}) of Gaussian form,
leaving the anharmonic term and a fraction of the harmonic term as part of the functions 
$f_1$ and $f_2$ in \eqref{eq:WUS}. 
This change is in part necessary because now we choose $\mu^2 < 0$ in our test cases, and this choice breaks down the 
positive definiteness of the matrix $C$ from the harmonic oscillator. 
Thus, we select a new covariance matrix $C^{\star}$ for the Gaussian samples, but we keep 
the sampling strategy for the $x_i$'s essentially unchanged as compared to the 
harmonic oscillator. 
The resulting weight functions in \ref{eq:WUS} are given by 
\begin{equation}
  \label{eq:weight_fns_anharm}
  f_1(\vz) = O( A^{\star} \Phi^{-1}(\vz) ) f_2(\vz) \; ,\quad 
 f_2(\vz) = e^{ - \sum_i 
   \left(a\left( \frac{\mu^2-\mu_{sim}^2}{2}\right) (A^{\star}_i \Phi^{-1}(\vz))^2 +  a \lambda (A^{\star}_i \Phi^{-1}(\vz))^4 \right)}\;,
\end{equation}
where $A^{\star}_i$ stands for the $i$-row of the factorization matrix $A^{\star}$ of $C^{\star}$, and $\mu_{sim}^2\ge 0$. 
As it can be seen from the resulting weight function $f_2$, we have chosen the simple strategy of calibrating the diagonal of the 
new covariance matrix $C^{\star}$ by the use of a parameter $\mu_{sim}^2\ge 0$.
Besides the requirement of positivity on $C^{\star}$, one is free in the choice 
of the parameter $\mu_{sim}$. We choose to follow the spirit of importance sampling by 
tuning $\mu_{sim}$ to a value that reduces the fluctuations of the weights $f_1$ and $f_2$ as much as possible. 
The samples based on the tuned parameter $\mu_{sim}$ lead us to observable averages with less variances and therefore smaller errors.
Nevertheless, it seems quite difficult to find an optimal criterion for the selection of the functions $f_1$, $f_2$ and the parameter  $\mu_{sim}$ which are leading to the best possible error behavior. At the moment we have to determine these quantities empirically and leave systematic investigations to the future.\\
Further, it is important to note that the PCA factorization during the generation of 
the Gaussian samples plays a mayor role for an efficient reduction of the effective dimension 
(see \cite{CAF97}) of the problem. For the parameters listed below, we estimated the effective 
dimensions in truncation sense $d_t$ of the functions \eqref{eq:weight_fns_anharm} to be close to $20$ 
(for a $99 \%$ variance concentration), for estimating the integrals described in \eqref{sec:Eff:Eff} 
with the dimension of the original system up to $d=1000$. Thus, we observe a drastic 
reduction of dimensionality. 

On the other hand, we found that the effective dimensions in truncation sense of the functions 
\eqref{eq:weight_fns_anharm} depend very strongly on the parameter 
$T = d a$, i.e. the physical time extent of the system, and seems not sensible to the real dimension $d$. 
We found that for small $T$-values, say $T < 0.2$, the selected parameters $\mu_{sim}$ and PCA
lead to a good effective dimension reduction, with $d_t$ close to $4$. In this case randomized QMC 
exhibited an $N^{-1}$ error scaling. 
The situation changed by increasing the $T$-values. For $T=1.5$ the 
effective dimensions $d_t$ significantly increased to be close to $20$. The exhibited error 
scaling was $N^{-\alpha}$ with $\alpha\approx 0.75$ for this case. 
Tests with values of  $T \geq 5$  indicate that the simulations become more and 
more difficult in the sense that one needs more and more samples to achieve 
the same accuracy of an observable as compared to estimates at $T=1.5$. Thus, in such situations the overlap of the 
sampling points with the functions $f_i$ in \ref{eq:WUS} seem to be too 
small to reduce the fluctuations sufficiently. 
It seems that for this problem there exists some kind of \textit{transition range} 
for the observed error scaling using randomized QMC in dependence of the time extent $T$, starting with 
a convergence rate $N^{-1}$ for $T$-values less than $0.2$ and decreasing to the poor convergence rate $N^{-1/2}$
for $T$-values higher than $5$.  
However, we are presently exploring a more general approach for selecting 
a good $T$-dependent matrix $C^{\star}$ (resp. $A^{\star}$ in \eqref{eq:weight_fns_anharm}) in the sampling procedure to improve
the situation for larger values of $T$. This question and the relation to the corresponding effective dimension deserves a detailed study, in particular when more realistic models are considered. However, such an investigation, although being very interesting, goes beyond the scope of the present paper.

\begin{figure}
\centering
\includegraphics[width =\textwidth]{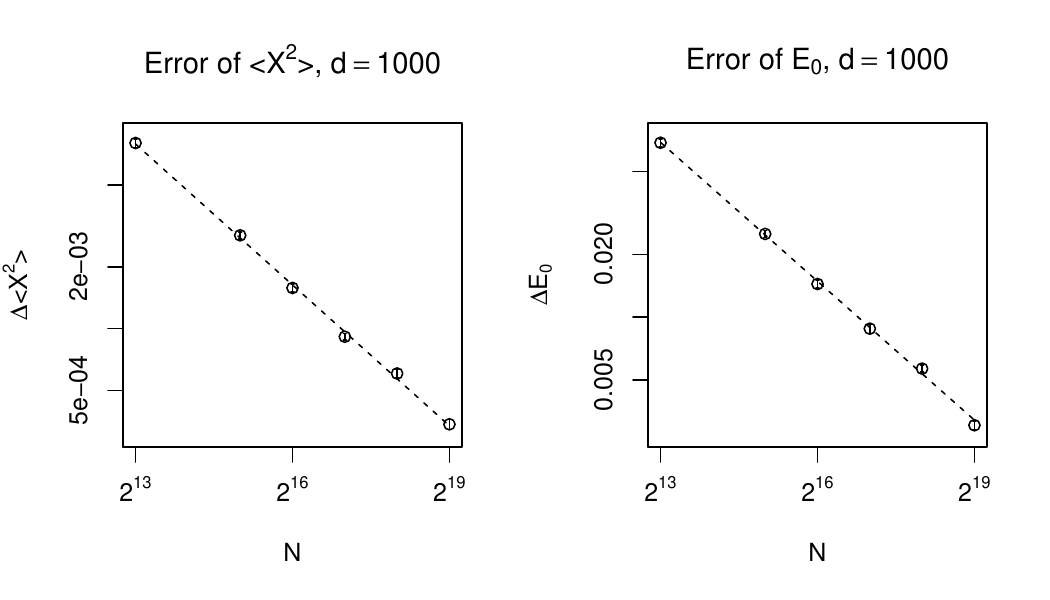}
\caption{We show the error of the observables $\langle X^2 \rangle$ and $ E_0$ 
as a function of the number of samples in a double logarithmic graph. 
The error of the error was obtained by performing 30 repetitions of the experiment 
with parameters chosen as $\lambda=1.0$, $\mu^2=-16$, $a=0.0015$ and $d=1000$.
For the sample generation randomly scrambled Sobol' (Rand. QMC) was used with $2^{13},2^{15},2^{16},2^{17},2^{18}$ 
and $2^{19}$ points. The dashed line shows the fit to the data points 
using a fit function $\log\left(  \Delta  \langle O \rangle  \right) \sim \log(C) + \alpha \log(N) $. The fitted 
exponents are $\alpha = -0.758(14)$ for $\langle X^2 \rangle$ and $\alpha = -0.737(13) $ for $ E_0 $, 
see also Table~\ref{tab:error_scaling_fit}. 
}
\label{sec:numex:fig:1}
\end{figure}

Nevertheless, for our numerical experiments, 
the parameters were set to $M_0 = 0.5$, $\lambda=1.0$, $\mu^2 = -16$.
In the two tests the lattice spacing $a$ was adjusted such, that $T$ was kept fixed. 
The tuned value of  $\mu^2_{sim}$ generally depends on all physical parameters of the 
system and in particular on $a$. Thus, we have to adjust also $\mu_{sim}$ when the lattice spacing $a$ is changed.
In particular, 
we set $a=0.015$ and $\mu^2_{sim} = 0.176$ for $d=100$, whereas for 
$d=1000$ $a=0.0015$ and $\mu^2_{sim}=0.2$ was chosen.
The error analysis of $\langle X^2 \rangle$ and $\langle X^4 \rangle$ was carried through 
in the same way as described for 
the harmonic oscillator test case discussed in the last 
subsection, \ref{ssec:HO}. 
We show in figure \ref{sec:numex:fig:1} the error of $\langle X^2\rangle$ and $E_0$ as a function
of the number of samples. In addition, we represent by the dashed line in 
figure \ref{sec:numex:fig:1} a 
fit to the data for the computed 
errors using the formula 
\begin{equation}
\log \left( \text{ Error } ( \langle O \rangle ) \right) = \log C + \alpha \log N \; ; \quad O = \{ x^2, x^4 , E_0 \} \; , 
\label{eq:fitfunction}
\end{equation}
with $C$ and $\alpha$ left as free parameters. 
From this analysis we can obtain a quantitative determination of the exponent of the error scaling.        
The results for the fit parameters are listed in Table \ref{tab:error_scaling_fit}.
\begin{table}
  \centering
  \begin{tabular}{|c|c|c|c|c|}
\hline\hline 
    & $O$ & $\alpha$ & $\log C $ &  $\chi^2 / \text{dof} $ \\
\hline\hline 
            &          &          &          &          \\
    $d=100$ & $ X^2  $ & -0.763(8) & 2.0(1) &  7.9 / 6   \\
            & $ X^4  $ & -0.758(8) & 4.0(1) & 13.2 / 6 \\
            & $E_0$ & -0.737(9) & 4.0(1) &  8.3 / 6 \\
            &          &          &          &          \\
\hline 
            &          &          &          &          \\
    $d=1000$ & $ X^2  $ & -0.758(14) & 2.0(2) & 5.0 / 4 \\
             & $ X^4  $ & -0.755(14) & 4.0(2) & 5.7 / 4 \\
             & $ E_0$ & -0.737(13) & 4.0(2) & 4.0 / 4 \\
\hline 
  \end{tabular}
\caption{Shown are the results for fit parameters of the 
error scaling for the observables considered, i.e. 
$ X^2 $ , $ X^4 $ and $ E_0 $, where the used 
fit function takes the form $\mathrm{error}=CN^\alpha$, see equation 
\ref{eq:fitfunction}. 
We also provide the $\chi^2$ values as well as the number of degrees of freedom 
in the fit (dof).}
  \label{tab:error_scaling_fit}
\end{table}

As can be inferred from Table~\ref{tab:error_scaling_fit}, in the case 
of the anharmonic oscillator the error scaling exponent is only $\alpha \approx 0.76$. 
However, this constitutes still a much improved error scaling compared 
to a Mc-MC simulation with a corresponding large gain in the number of
required samples to reach a desired accuracy. 
Moreover, the value of $\alpha$ is consistent for all observables 
considered here and independent from the dimension of the 
problem $d$, a finding which is clearly encouraging. 

We finally mention that 
the resulting 
estimates of the ground state energy for $T=1.5$ matches in at least two significant digits with the theoretical 
value, $E_0 = 3.863$, calculated in \cite{Blank79}, namely $\hat{E}_0 = 3.857 \pm 0.004 $ for 
$d=100$ and $\hat{E}_0 = 3.862\pm0.004$ for $d=1000$.

\section{Concluding Remarks}

In this article we have performed a first application of QMC methods 
to Euclidean 
lattice models. The goal was to see, whether QMC 
algorithms provide also in the case of non-Gaussian systems an improved 
error scaling behavior with respect to Markov-chain Monte Carlo methods. 
As a prototype system, we have considered the quantum mechanical oscillator
discretized on a Euclidean time lattice, 
both in its harmonic (Gaussian) form as well as adding a non-Gaussian quartic 
term (anharmonic oscillator). 
For the harmonic oscillator we found a large-$N$ ($N$ being the number
of sample points) immproved error behavior, i.e.  
$\sim N^{-1}$ for (randomized) QMC and $\sim N^{-1/2}$ for Mc-MC.

The main result of our investigation is that 
also for the anharmonic oscillator, which 
is a
non-Gaussian 
problem, the QMC approach leads to a significant 
improvement of the error scaling $N^\alpha$ with $\alpha \approx -0.76$, see 
Table~\ref{tab:error_scaling_fit} for the exact values of $\alpha$ for different observables 
and different physical situations.  

Further, 
we found 
that the accessible range of $T=1.5$ values gives already estimates of the 
ground state energy, compatible (within errors) with the theoretical prediction 
(valid in the limit $T\rightarrow \infty$ and $a\rightarrow 0$).
For the case that the improved error scaling and the mild dependence 
on the lattice spacing $a$ found here will also be present in more 
elaborate models, QMC methods have the potential to become very valuable in the future.
On the other hand we observed 
that the applicability of the WUS (reweighting) approach 
seems to be limited by the 
physical time extent $T=d a$ of the system. 
For values of $T\leq 1.5$ the error falls below the percent
level within the investigated number of samples. For increasing
values of $T$ the error is continuously growing and at $T=5$ we can only
obtain a relative error of $\approx 15 \%$ for $2^{19}$ samples. For
larger $T$ values we expect the error to become even larger leading 
eventually
for very large $T$ to a situation where a meaningful evaluation of the
considered quantities is not possible anymore. This behavior and the relation to the effective dimension of the problem clearly needs an understanding and a dedicated investigation in the future, in particular, when more realistic models are considered. 

It is clear that the here considered quantum mechanical systems are rather
simple models and still a long way has to be gone, if generic quantum field 
theories, especially gauge theories are to be studied. Nevertheless, it is 
very reassuring that we find an improved error scaling behavior in the 
case of a quartic potential and hence a non-Gaussian system. 
This promising result is certainly a strong motivation for studying further
QMC methods in lattice field theories and statistical mechanics.

\section*{Acknowledgment}
The presented research cooperation has been supported by the Center 
of Computational Sciences Adlershof (CCSA), Berlin, Germany. 
The authors wish to express their gratitude to Alan Genz (Washington State University) 
and Frances Kuo (University of New South Wales, Sydney) for inspiring comments and 
conversations, which helped to develop the work in this article. Frances Kuo 
collaborated with us during her visit to the Humboldt-University Berlin in 2011. 
A.N., K.J. and M.M.-P. acknowledge financial support by the DFG-funded corroborative 
research center SFB/TR9. 
K. J. was supported in part by the Cyprus Research Promotion
Foundation under contract $\Pi$PO$\Sigma$E$\Lambda$KY$\Sigma$H/EM$\Pi$EIPO$\Sigma$/0311/16.
We would like to thank specially the reviewers for their suggestions and comments which
help us to improve the article.

\section{Appendix: C++ programme for the QMC simulation of the (an)harmonic oscillator}
We provide a C++ programme for the QMC simulation of the (an)harmonic oscillator at the following URL \url{http://arxiv.org/format/1302.6419v3} (``Source''). In the following we give a short description for the usage of the programme.
The programme implements the suggested algorithm for the evaluation of the path integral of the harmonic and anharmonic oscillator in the QMC approach, except that the present implementation applies random digital shifts instead of random scramblings to generate different sobol sequences. Eventually, both methods should lead to very similar results.\\
Copyright information may be obtained from the file ``README'' in the package.
The package is equipped with a standard ``Makefile'' and a cmake input file ``CMakeList.txt''.
\subsection{Prerequisites}
The only external dependency is the FFTW library version 3 which can be obtained from \url{http://www.fftw.org}. The FFTW library offers an efficient implementation of the Hartley transform.\\
If the  library is installed in a non-standard path of your PC, you can adjust the variable ``FFTWDIR'' in the very beginning of the make file or via the environment variable ``FFTWDIR'' when using ``CMakeLists.txt''.

\subsection{Building}
Simply run '\verb+make+' when using ``Makefile'' or use '\verb+ccmake <path to package source>+' in an empty directory followed by a '\verb+make+'.

\subsection{Parameters}
Having built the executable '\verb+qmc_quartic_reweight+' you can run the programme, preferably in a new empty directory, and may pass the following parameters:
\begin{center}
\begin{tabular}{l|r}
{\bf Parameter }      &  {\bf Meaning } \\
\hline 
\verb+-N <Integer>+ & number of dimensions \\
\verb+-k <Integer>+ & number of samples per estimation \\
\verb+-c <Integer>+ & number of configurations written out to a file \\
\verb+-S <Integer>+ & max. time separation for correlator \\
\verb+-a <Float>+   & lattice spacing $a$ \\
\verb+-M <Float>+   & particle mass $M_0$\\
\verb+-m <Float>+   & $\mu^2_{sim}$ \\
\verb+-u <Float>+   & $\mu^2$ \\
\verb+-l <Float>+   & $\lambda$ \\
\verb+-Q <Path to file>+ & file containing directions numbers
\end{tabular} .
\end{center}
Files with direction numbers can be obtained from Frances Kuo's page \url{http://web.maths.unsw.edu.au/~fkuo/sobol/index.html}.
The programme produces 10 estimations, each with the given number of samples. Output is written to files of the form \\
\verb+<prefix>_s1_N<# dimensions>_a<a>_M0<M0>_musq<mu^2_sim>_l<lambda>_J0.000000.csv+.
The result of each estimation is stored in the file with the prefix ``obs\_macro''. The first column contains the estimated value of $\langle x^2 \rangle $ and the second column contains $\langle x^4 \rangle$. Successive runs of the programme  with the same parameters will append 10 more estimates. Correspondingly, 30 runs  should suffice to produce the statistics we used in this work.

\begin{small}
\bibliography{Articles}
\bibliographystyle{unsrt}
\end{small}

\end{document}